# Trends in porous media laboratory imaging and open science practices

Na Liu[1*], Jakub Wiktor Both[2], Geir Ersland[1], Jan Martin Nordbotten[2], Martin Fernø[1,3]

[1]Department of Physics and Technology, University of Bergen, Bergen, Norway

[2]Department of Mathematics, University of Bergen, Bergen, Norway

[3] Norwegian Research Centre AS – NORCE, Nygårdsgaten 112, 5008, Bergen, Norway

*Corresponding author: Na Liu, Na.Liu@uib.no

**Abstract**: Understanding processes in porous media is fundamental to a broad spectrum of environmental, energy, and geoscience applications. These processes include multiphase fluid transport, interfacial dynamics, reactive transformations, and interactions with solids or microbial components, all governed by wettability, capillarity, and reactive transport at fluid–fluid and fluid–solid interfaces. Laboratory-based multiscale imaging provides critical insights into these phenomena, enabling direct visualization and quantitative characterization from the nanometer to meter scale. It is essential for advancing predictive models and optimizing the design of subsurface and engineered porous systems. This review presents an integrated overview of imaging techniques relevant to porous media research, emphasizing the type of information each method can provide, their applicability to porous media systems, and their inherent limitations. We highlight how imaging data are combined with quantitative analyses and modeling to bridge pore-scale mechanisms with continuum-scale behavior, and we critically discuss current challenges such as limited spatio-temporal resolution, sample representativity, and restricted data accessibility. We conduct an in-depth analysis on open-science trends in experimental and computational porous media research and find that, while open-access publishing has become widespread, the availability of imaging data and analysis code remains limited, often restricted to 'upon request'. Finally, we underscore the importance of open sharing of imaging datasets to enable reproducibility, foster cross-disciplinary integration, and support the development of robust predictive frameworks for porous media systems.

**Keywords**: Advanced imaging technology, porous media, fluid-fluid interfaces, multiphase flow, image analysis, open data trends.

**Highlights:**

- Laboratory imaging enables multiscale visualization of fluid and pore structures.
- 2D and 3D imaging provide insights into flow, trapping, and reactive transport.
- Multiscale imaging and segmentation enable quantitative porous media analysis.
- Open data access and sharing are vital for reproducibility and robust modeling.

**Graphical abstract**

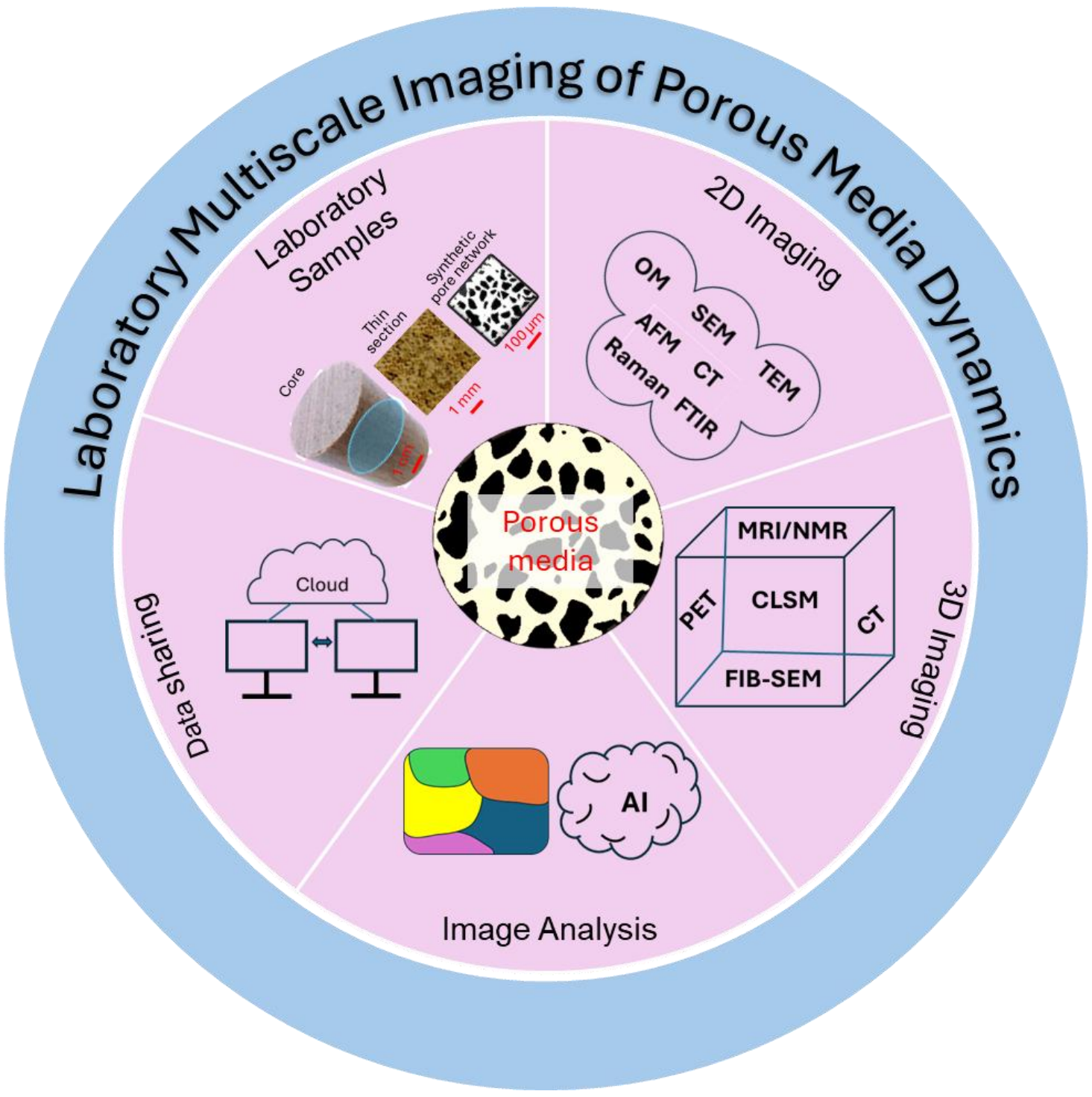

## 1. Introduction

The behavior of fluids in porous media is governed by complex physical processes that occur across multiple scales, including fluid displacement, phase distribution, capillary flow, and interactions with solid surfaces and microbial communities. These processes underpin a wide range of applications, including petroleum and hydrocarbon recovery, subsurface geoengineering, groundwater management, soil remediation, geothermal energy and gas storage in geological formations. The heterogeneous and multiscale nature of pore networks—characterized by variations in pore size, connectivity, surface roughness, mineralogy and surface properties—leads to fluid behavior that cannot be reliably inferred from bulk properties alone.

Advances in high-resolution imaging techniques have revolutionized the ability to observe and quantitatively characterize porous media processes across scales. Laboratory imaging provides detailed structural and functional information, including pore geometry, connectivity, porosity, fluid distributions, multiphase flow, capillary trapping, dissolution, and interfacial dynamics [1]. Beyond capturing fluid movement, these techniques allow precise characterization of pore structures—including porosity, connectivity, tortuosity, and heterogeneity—providing essential data for validating theoretical and computational models and linking microscopic processes to macroscopic transport properties [2, 3]. In addition, imaging enables controlled laboratory experiments that probe processes that cannot accurately be assessed at the field site, providing critical insights into multiphase flow, reactive transport, and fluid–rock–microbial interactions.

Despite extensive studies on imaging techniques, a comprehensive review that integrates both two-dimensional and three-dimensional laboratory methods compares their capabilities, and discusses their limitations and applications is currently lacking in the porous media literature. This review therefore aims to provide a unified perspective on laboratory imaging, focusing not on the operational details of the instruments, but on the type of information each technique can reveal about subsurface porous systems. We highlight how different methods complement one another, where their strengths lie, and what constraints must be considered when applying them to porous media. Importantly, we emphasize the importance of making experimental data and original images openly accessible, as these are critical for model validation, calibration of computational analyses, and reproducibility—yet are often absent in previous studies. The goal is to offer researchers and practitioners a structured framework for selecting appropriate imaging methods to answer specific scientific or engineering questions, interpreting the resulting datasets, and integrating insights into predictive models of porous media processes.

This review is structured as follows: Sections 2 and 3 introduce two- and three-dimensional imaging techniques, detailing their principles, strengths, limitations, and complementarity across scales. Section 4 focuses on laboratory applications, highlighting pore-scale dynamics, multiphase flow, reactive transport, and emerging multimodal approaches. Section 5 discusses quantitative image analysis and open science, covering workflows, segmentation, upscaling, machine learning, and trends in data and code availability. Finally, Section 6 presents perspectives and conclusions, emphasizing the integration of advanced imaging, modeling, and open-access frameworks to advance predictive understanding and cross-disciplinary collaboration in porous media research.

## 2. Two-dimensional imaging techniques

Understanding single- and multiphase pore-scale processes in porous media requires visualization of internal structures, pore geometry, fluid-fluid and fluid–solid interactions. Two-dimensional (2D) imaging techniques are widely used for this purpose, providing high-resolution information on pore morphology, connectivity, and multiphase flow dynamics. Compared to three-dimensional (3D) approaches, 2D methods are typically more cost-effective and computationally efficient, which enables studies at very different scales and resolutions. For example, quasi-2D concepts such as the meter-scale FluidFlower allow experiments to be conducted over much larger spatial dimensions, thereby capturing Darcy-scale impacts of pore-scale processes [4]. At the other extreme, electron microscopy provides nanometer-scale pore characterization, enabling observation of pore morphology and surface roughness (e.g. [5, 6]). This multiscale capability makes 2D imaging a powerful approach for linking pore-scale physics with continuum-scale behavior, and for studying capillary forces, wettability, fluid displacement, reactive transport and other transport phenomena, forming the basis for quantitative modeling and upscaling laboratory observations to reservoir scale. In the following sections, we describe different 2D imaging techniques, highlighting their capabilities, limitations, and applications in porous media research.

### *2.1 Optical and Light-Based Imaging Techniques*

Photography provides a straightforward and versatile approach for capturing 2D images of porous media at larger scales, typically from centimeters to decimeters, where microscopy is impractical. It has been widely applied in studies using transparent synthetic media, such as glass bead packs [7], microfluidic setups [8, 9] and FluidFlower rigs [10], to visualize pore networks, fluid distributions, and coupled free flow–porous media experiments. Modern experiments often use charge-coupled device (**CCD**) cameras, which offer high spatial resolution, excellent light sensitivity, and the ability to record dynamic processes with high temporal resolution. CCD cameras are particularly useful for time-lapse or high-speed imaging of dynamic processes such as fluid displacement [11], convective mixing [12], or biofilm development [13], bridging the gap between pore-scale observations and macroscopic, effective-scale phenomena.

Optical microscopy (**OM**) is one of the most widely applied approaches for visualizing porous media, owing to its simplicity, accessibility, and ability to directly image thin sections or polished surfaces. Using visible light, OM achieves a spatial resolution of approximately 200 nm, constrained by the diffraction limit. In porous media research, OM is commonly employed to study pore geometry, mineralogy, and grain textures using thin sections prepared by standard petrographic techniques [14]. Both transmitted and reflected light modes can be applied: transmitted light allows the visualization of pore structure, mineral inclusions, and cementation in thin sections, whereas reflected light is useful for opaque minerals and surface features [15].

Polarized Light Microscopy (**PLM**) extends the capabilities of OM by introducing crossed polarizers, which highlight optical anisotropy and crystalline structures [16]. This technique is particularly effective for mineral identification and for characterizing features such as grain boundaries [17], cement phases, and preferred mineral orientations [18], which are important for understanding mechanical and transport properties of the rock matrix.

Fluorescence Microscopy (**FM**) enables selective visualization of organic matter, tracers, or microbial biofilms by exploiting fluorescence signals emitted by dyes or naturally fluorescent compounds. In porous media studies, FM has been used to trace fluid pathways [19], detect biofilm growth in pore networks [20], and monitor the distribution of organic coatings that affect wettability and flow behavior [21].

Confocal Laser Scanning Microscopy (**CLSM**) provides higher resolution than OM and depth control by using a laser scanning system with a pinhole aperture that rejects out-of-focus light. This enables optical sectioning of thick, hydrated samples and the reconstruction of 3D structures from stacked 2D images [22]. CLSM is particularly useful for studying biofilm colonization [23] and multiphase fluid distributions *in situ* [22], under conditions that are difficult to access with conventional OM or scanning electron microscopy [24]. Its ability to non-destructively image wet and soft systems make it a valuable complement to electron microscopy and X-ray techniques. In addition, CLSM has proven more sensitive than the water drop penetration time test for detecting the effects of wettability, which diminish with increasing water content [25].

### *2.2 Electron-based Techniques*

Scanning Electron Microscopy (**SEM**) is a widely used technique for high-resolution, 2D imaging of porous media surfaces, offering detailed insights into pore morphology, grain surfaces, fracture networks, and mineral textures [26, 27]. SEM operates by scanning a focused electron beam over the sample, which generates secondary electrons detected to form high-resolution gray-scale images with nanometer-scale precision. To preserve microstructure and prevent charging, samples are typically fixed, dehydrated, and coated with a thin conductive layer before imaging. Modern SEM instruments can achieve magnifications up to 300,000×, and some advanced models reach 1,000,000×—far exceeding the magnification range of optical microscopes (typically 400–1000×) [27]. Its high magnification and resolution make SEM particularly useful for analyzing surface morphology, microcracks, particle deposition, and biofilm colonization [28]. SEM can be combined with Energy Dispersive X-ray Spectroscopy (EDS) to obtain qualitative and semi-quantitative elemental maps, allowing simultaneous structural and chemical characterization [28]. This combination reveals mineralogical changes and their impact on porous media, such as caprock porosity enhancement and microcrack formation due to calcite or dolomite dissolution [29]. For nanoscale structures, focused ion beam-scanning electron microscopy (**FIB-SEM**) nanotomography is the method of choice, resolving pore and particle diameters below 100 nm [30].

Backscattered Electron Imaging **(BSE-SEM)** provides compositional contrast in porous media by producing image intensity proportional to atomic number, with heavier elements appearing brighter and lighter elements darker. This enables differentiation of minerals and phases [31], visualization of grain boundaries, microcracks, mineral coatings and precipitates on pore surfaces [32]. When combined with EDS, BSE-SEM allows mapping of elemental distributions and semi-quantitative chemical analysis [33].

Environmental SEM (**ESEM**) further expands SEM capabilities by allowing imaging of samples in the presence of fluids (e.g., oil or brine) without freezing [34]. ESEM enables observation under controlled parameters, including atmosphere (air, vapor, nitrogen, carbon dioxide), temperature (−20 °C to 1000 °C), hydration rates, and reagent injection, making it suitable for studying wettability [34], fluid distribution [35], and *in situ* interactions [35, 36].

Transmission Electron Microscopy **(TEM)** enables imaging at the nanometer and sub-nanometer scale by transmitting electrons through ultrathin samples. TEM provides detailed information on nanoparticle transport [37], mineral crystallography, and nanostructures [5] that are not resolvable by SEM. Although its application in porous media is less common, TEM is particularly useful for characterizing fine-grained materials such as shales [38], and clays [39], as well as for analyzing particle–mineral interactions at the nanoscale [40]. Using TEM, nanoparticles attached to pore walls, adsorbed on pore surfaces, or captured at pore throats can be directly observed, providing insight into mechanisms of nanoparticle transport and retention within porous media [41].

### *2.3 Surface Techniques*

Atomic Force Microscopy (**AFM**) is a versatile technique for imaging porous media surfaces at nanometer resolution, providing detailed topographical maps and quantitative measurements of surface roughness [42, 43], pore geometry [44, 45], and mechanical properties such as stiffness or adhesion [46]. Unlike SEM, AFM does not require conductive coatings or vacuum conditions, allowing imaging of hydrated samples or soft materials such as biofilms and polymeric pore linings [47]. AFM can also probe nanoscale interactions (<0.1 nm) between fluids and pore surfaces, offering insights into wetting, adsorption, and microbial colonization processes that directly influence transport and reaction dynamics in porous systems [6]. Colloidal–AFM has been used, for example, to characterize interactions between colloids and mineral surfaces [44, 45].

Scanning Tunneling Microscopy **(STM)** measures the quantum tunneling current between a conductive tip and a conductive or semiconductive surface, enabling atomic-resolution imaging of surface electron density, atomic arrangements, and defects. However, STM requires electrically conductive samples, which limits its application in many geoscience and porous media studies. For example, STM has been used to image the surfaces of small metal particles formed in anodic oxide pores, providing atomic-scale structural information [48].

### *2.4 X-ray-based 2D Methods*

X-ray Microtomography **(X-Ray CT)** – 2D slices is a non-destructive imaging technique that generates internal cross-sectional images of porous media by reconstructing 2D slices from multiple X-ray projections [49]. In laboratory micro-CT (**μCT**), X-rays interact with matter primarily through inelastic scattering and photoelectric absorption, with beam attenuation depending on the geometry, density, and atomic composition of the sample. Laboratory μCT typically achieves a resolution of ~0.5–5 μm, while synchrotron-based systems can reach ~30 nm, enabling detailed imaging of pore shape, size, connectivity, and mineral distribution [50]. Under X-ray irradiation, high-density minerals such as quartz, calcite, or dolomite exhibit strong X-ray attenuation and are readily resolved in μCT images. This allows clear visualization of grain boundaries, microcracks, and mineral heterogeneities. In contrast, liquid and gases (e.g., water, oil, carbon dioxide) have lower X-ray attenuation than the solid matrix, enabling fluid-saturated pores to be distinguished from surrounding grains, particularly when contrast agents are used to enhance detection. Gaseous phases, including air, carbon dioxide or hydrogen, exhibit very low X-ray attenuation and appear as dark regions, facilitating the identification of gas-filled pores or bubbles [51, 52]. CT imaging relies on sufficient density contrasts which are often not fulfilled in low-porosity media.

While individual 2D μCT slices provide planar views useful for quantitative analysis of porosity, fracture patterns, and local heterogeneity, they inherently lack volumetric information. Accurate assessment of 3D pore connectivity and network properties requires stacking multiple slices into a full 3D reconstruction. This step is essential for evaluating transport properties, fluid flow, and reactive processes but can be limited by sample size, imaging resolution, and computational constraints. Despite these limitations, μCT remains a powerful tool for rapid, non-destructive characterization of porous media, bridging the gap between microscale structure and bulk material behavior.

### *2.5 Spectroscopic imaging*

**Raman** and Fourier Transform Infrared (**FTIR**) microscopy are non-destructive techniques that provide spatially resolved chemical information of porous media surfaces. Raman microscopy measures the inelastic scattering of monochromatic laser light, which reflects molecular vibrations and chemical composition, while FTIR detects the absorption of infrared light to identify functional groups and molecular bonds. By scanning the sample surface, both methods can generate 2D chemical maps, allowing visualization of mineral compositions [53], component concentration [54], or biofilm distribution within pores [55]. These spectroscopic imaging techniques complement structural methods such as OM, SEM, or μCT. Raman microscopy, in particular, can be combined with microfluidic setups to enable *in situ* chemical analysis of fluids, minerals, and biofilms under controlled flow conditions, linking chemical composition to dynamic pore-scale processes [53, 56]. FTIR microscopy similarly allows mapping of chemical heterogeneity and functional groups at the surface [57]. Together, these techniques provide insights into mineralogical variability, organic–inorganic interactions, fluid–mineral reactions, and microbial colonization, helping to elucidate transport, reaction, and biofilm development in porous systems.

### *2.6 Evaluation of two-dimensional imaging techniques*

2D imaging techniques can provide essential insights into pore-scale heterogeneity, mineral composition, and biogeochemical processes in porous media. **Table 1** summarizes the main 2D imaging methods, detailing their strengths, limitations, destructive potential, and typical combinations, and illustrates how these complementary techniques span multiple spatial and temporal scales.

Optical imaging methods excel at capturing large areas and dynamic processes in real time, despite their lower resolution compared to electron or scanning probe techniques. They are increasingly combined with laboratory synthetic porous media, such as microfluidic devices and glass models, to monitor fluid dynamics under controlled conditions [58]. Applications of these methods in dynamic flow studies will be discussed further in Section 4.

Electron-based microscopy provides high-resolution imaging of surface morphology, microcracks, and mineral textures. However, these techniques are primarily static, capturing snapshots of the sample at a specific time. Temporal changes in pore structure, wettability, or fluid distribution cannot be observed directly, and sample preparation—such as drying, coating, or exposure to vacuum—may alter the natural state or introduce artifacts. Even with ESEM, which permits imaging under fluid-saturated or controlled atmospheric conditions, *in situ* conditions are only partially reproduced. Consequently, SEM is often complemented with techniques such as μCT, X-ray tomography, or *in situ* optical and synchrotron-based imaging to enable 3D and time-resolved visualization of dynamic processes.

Surface-probe techniques, such as AFM, provide nanometer-scale topography and quantitative mechanical measurements, making them a powerful complement to SEM. AFM is particularly effective for studying surface roughness, nanopore distribution, and local mechanical heterogeneity, and it can operate in liquid environments to probe microbe–mineral and fluid–mineral interactions. Its main limitations are the small, scanned areas and the inability to capture dynamic, time-dependent processes at larger scales.

Spectroscopic imaging techniques provide non-destructive chemical mapping at the pore scale. Complementing structural imaging, they reveal mineral composition, functional groups, or biofilm distribution, and can be coupled with microfluidics for *in situ* analysis. While their spatial resolution is lower than electron or probe-based methods, they add a valuable chemical perspective and are often combined with optical, electron, and X-ray techniques for a multiscale understanding of porous media.

**Table 1.** Summary of two-dimensional imaging techniques.

| Imaging technique | Strengths | Weaknesses | Destructive | Often combined with |
|---|---|---|---|---|
| Light-based microscopy | Large-area coverage, dynamic processes, pore-scale heterogeneity, mineral composition, biogeochemical processes, multi-colored tracers | Limited spatial resolution (except CLSM); may not resolve nanoscale features; only transparent media, | No | Microfluidic devices, glass model, FluidFlower setups, SEM, μCT |
| Electron-based microscopy | High resolution; surface morphology; chemical mapping with EDS | Mostly static; sample prep can alter structure; small field of view | Often (coating, vacuum, thin sections) | AFM, μCT, spectroscopic imaging |
| Surface-based microscopy | Nanometer-scale topography; mechanical property mapping | Small scan area; mostly static; STM requires conductive samples | Usually | SEM, quantitative mechanical measurements |
| X-Ray | Rapid, non-destructive internal structure; visualize pore connectivity and mineral distribution | Low resolution; density contrast required | No | SEM, spectroscopy |
| Spectroscopy | Chemical composition mapping; in situ fluid/solid/biofilm analysis | Limited spatial resolution (compared to SEM/AFM); slower scans for large areas | No | OM, CLSM, microfluidics, µCT |

Together, these imaging methods provide complementary insights across scales. As illustrated in **Figure 1**, PLM highlights mineralogical variations in a Berea sandstone thin section, while X-ray µCT reconstructs sub-millimeter internal grain and pore structures. High-resolution SEM reveals detailed surface morphology, and AFM maps surface topography at sub-nanometer resolution. By spanning nanometer to millimeter scales, these techniques collectively offer a comprehensive, multiscale view of rock–fluid–gas–microbe interactions in porous media.

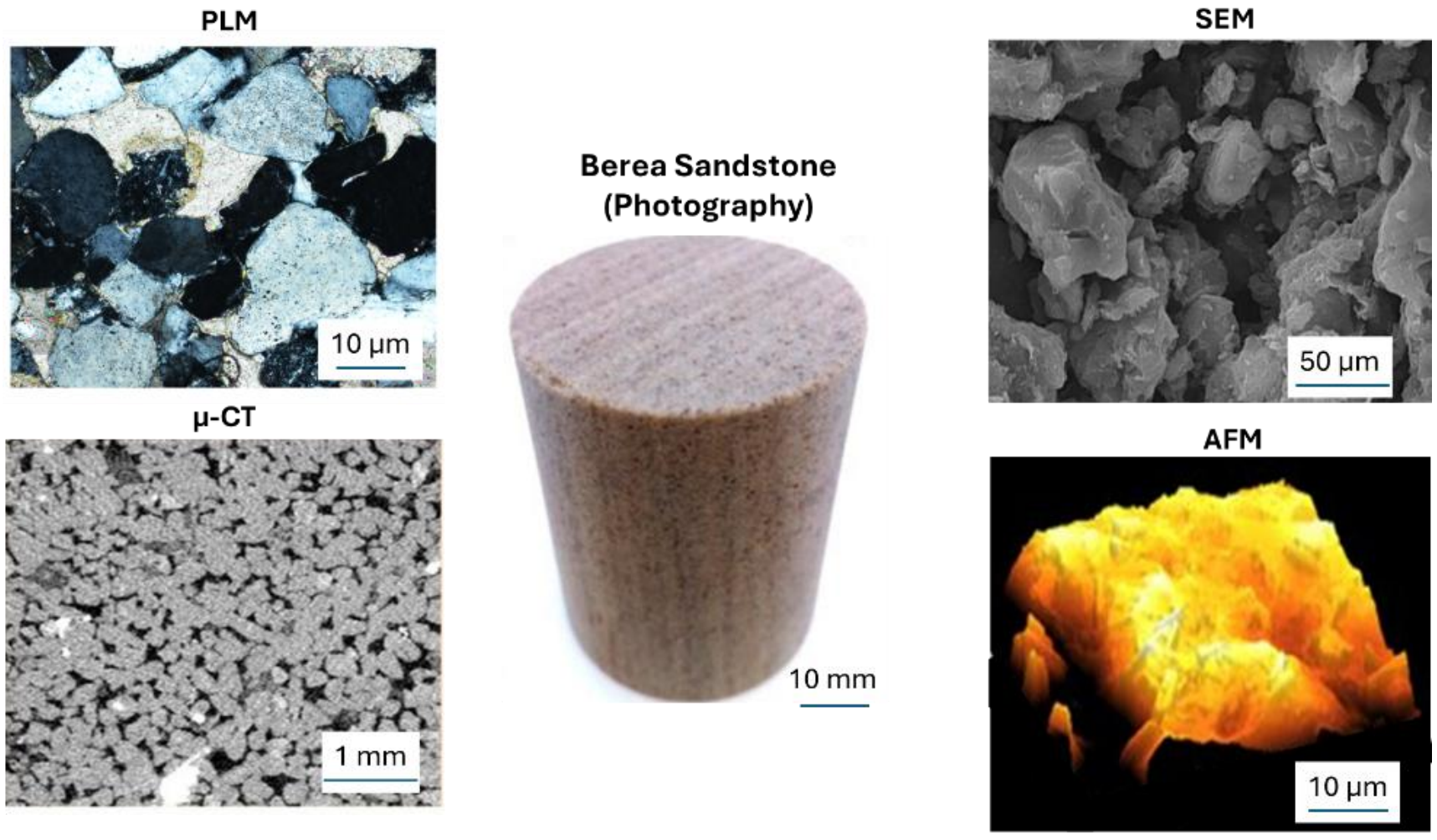

***Figure 1**. Example of multi-scale, multi-modal characterization of a natural Berea sandstone sample (center) using optical microscopy (OM), X-ray µCT, scanning electron microscopy (SEM), and atomic force microscopy (AFM). Upper left: OM image of a Berea thin section obtained with a Nikon polarizing microscope at 200× magnification with crossed polarizers. Reproduced from [59] with permission; Lower left: X-ray µCT image reveal the sub-mm internal grain and pore structure of the sample. Reproduced with permission from [60]; High-resolution SEM image of the surface morphology. Reproduced with permission from [61]; Sub-nanometer resolution maps of surface topography obtained by AFM. Reproduced with permission from [62].*

## 3. Three-dimensional imaging techniques

3D imaging techniques provide volumetric insight across multiple scales into porous media, from the pore scale (micrometers) to sample or core scale (millimeters to centimeters). These methods enable quantitative analysis of pore geometry, connectivity, fluid distributions, and particle transport, linking microscale structural features to macroscopic transport, reaction, and microbial processes. Techniques such as laboratory µCT and magnetic resonance imaging (MRI) can capture larger-scale or macroscopic events, such as multiphase flow patterns, convective mixing, and bulk transport in centimeter-scale samples, while high-resolution methods like CLSM and FIB-SEM are restricted to pore- to sub-micron scales. 3D imaging technologies can be broadly categorized into reconstructed 3D methods, which build volumetric data from multiple 2D slices, and direct 3D methods, which inherently capture volumetric information.

### 3.1 Reconstructed 3D Imaging from 2D Slices

Many high-resolution imaging methods, such as CT, CLSM and FIB-SEM nanotomography, generate volumetric data by stacking multiple 2D images. These approaches provide high-resolution structural information but require computational reconstruction and alignment of individual slices.

*Computer tomography*

The principles of X-ray **CT** imaging have been discussed extensively in Section 2. In µCT, a sample is rotated while multiple 2D radiographs are acquired from different angles, and these slices are computationally reconstructed into a 3D volume, revealing the internal architecture of the sample without destruction. From 3D µCT images, both structural and functional information can be extracted, allowing for quantitative characterization of pore geometry, including size distribution, shape, volume, surface area, and connectivity, as well as segmentation into multiple phases such as pores and solid grains [63]. Differences in X-ray attenuation enable identification of mineral phases, grain boundaries, and microstructural defects such as microcracks and heterogeneities, which influence permeability, particle retention, and mechanical behavior [64].

**µCT** can also capture real-time fluid and interface distributions in 3D, allowing measurement of in situ contact angles, interfacial tension, capillary pressures, and residual fluid distributions [65]. As a non-destructive technique, it enables time-resolved studies of multiphase flow without disturbing the system, providing insights into microscopic displacement mechanisms critical for secondary and tertiary oil recovery [66, 67]. Time-resolved µCT also allows monitoring of dynamic processes such as particle transport, fluid migration, and chemical reactions, providing temporal information within the pore network [65, 68, 69]. These capabilities have been applied in a range of case studies, including permeability estimation of porous media [64], structural analysis of natural gas hydrate [68], and pore-scale imaging of gas displacement and trapping [66, 67, 70, 71]. Combined with computational modeling, µCT data can predict flow-induced shear stress, heat and mass transfer, and mineral reactions [72-75].

**Neutron CT** is a complementary non-destructive 3D imaging technique that uses neutrons instead of X-rays, interacting with atomic nuclei rather than electron clouds. This provides strong contrast for light elements such as hydrogen, water, or lithium, while many metals remain nearly transparent. As a result, neutron CT is particularly effective for visualizing fluids, hydrogen-rich phases, and water distribution in dense or opaque porous samples. Similar to µCT, it involves rotating the sample in a neutron beam to acquire 2D radiographs that are reconstructed into a 3D volumetric image, revealing pore structure, fluid distribution, and transport pathways [76, 77].

*Confocal Laser Scanning Microscopy*

**CLSM** generates 3D high-resolution images by scanning a focused laser beam across the sample and acquiring multiple 2D optical sections at different depths, which are then stacked computationally to reconstruct the full 3D architecture. CLSM has been widely used in the biological community over the past decade [78], and has been proved effective as an optical imaging technique in the field of geomaterials [79]. It has been applied to examine pore networks in sandstone reservoir rocks [80] and to characterize porosity in hardened concrete [81]. Fluorescent labeling of fluids, particles, or biological components enhances phase contrast, enabling segmentation of pore spaces, liquid phases, and solid surfaces [82]. The method also allows quantification of parameters such as local fluid saturation, flow pathways, and particle deposition patterns [24, 25]. CLSM has seen limited use for 3D imaging of geomaterials due to its shallow penetration depth (<500 µm) [80, 83]. It enables high-resolution, non-destructive visualization of pore–grain geometries and can be combined with optical setups, such as microfluidics or glass models, for pore-scale studies of multiphase flow in porous media [84].

*Focused Ion Beam–Scanning Electron Microscopy*

**FIB-SEM** tomography is a high-resolution 3D imaging technique that combines serial sectioning with conventional SEM imaging. In FIB-SEM, a focused ion beam mills successive thin slices from a sample, and an SEM image is captured after each milling step, producing a stack of 2D images corresponding to successive planes of the material. Computational reconstruction of these slices generates a 3D representation of the internal structure, enabling detailed visualization of pore networks, grain arrangements, and microstructural heterogeneities [85].

The method provides nanometer-scale resolution, with slice thicknesses and pixel sizes typically in the order of tens of nanometers, capturing features at the mesoscopic scale [86, 87]. This allows quantitative characterization of pore size distributions [88], connectivity [89], mineral phases [89], and microstructural defects such as microcracks or cementation [87]. Unlike conventional SEM, which is limited to surface imaging, FIB-SEM tomography delivers true 3D information at the nanoscale, bridging pore-scale structure and macroscopic transport behavior. Applications in geomaterials include sandstones, claystones, and carbonates, where FIB-SEM has been used to assess the impact of mineralogy, cementation, and clay content on 3D pore networks [1, 90, 91]. Coupled with image analysis and computational modeling, FIB-SEM data can link mesoscale structural features to macroscopic properties such as permeability, transport pathways, and reactive surface area.

### *3.2 Direct Three-dimensional Imaging Methods*

Direct 3D imaging techniques capture the internal structure of porous materials without relying on reconstruction from 2D slices, thereby reducing alignment errors and artifacts associated with serial sectioning. These methods provide volumetric data of pore networks, solid matrices, and fluid distributions, often at high resolution and with minimal assumptions. Common approaches include nuclear imaging–based methods and neutron tomography, each offering trade-offs between sample volume, spatial resolution, temporal resolution, and material contrast. These methods are particularly suited for non-destructive imaging of fluid saturation, flow dynamics, and transport processes within opaque porous media.

*Nuclear Magnetic Resonance and Magnetic Resonance Imaging*

Nuclear Magnetic Resonance (**NMR**) utilizes the magnetic properties of atomic nuclei (typically hydrogen) to provide information of pore size distributions, fluid content, wettability, phase transitions, and transport processes in porous media. Magnetic Resonance Imaging (**MRI**) extends NMR by introducing magnetic field gradients to spatially encode signals, enabling non-destructive 1D, 2D, and 3D imaging. MRI can deliver 1D saturation profiles in the order of seconds, high-resolution 2D images and full 3D images within minutes. MRI has emerged from a medical diagnostic tool into a versatile technique for investigating a wider range of porous media from fuel cells [92] to building materials [93] and porous rocks [94]. In subsurface applications, MRI is most commonly used for core sample characterization, to acquire pore size distributions and fluid saturations, but also more advanced studies of displacement processes [94], formation damage [95],enhanced recovery methods [94, 96, 97], geological CO2 storage and gas hydrates [98]. MRI of sediments typically does not resolve the pore space due to time constraints, as achieving sufficient resolution requires prolonged acquisition times. Although successful demonstrations have been reported [99], the technique remains limited in practical applications. However, a significant advancement enabling spatial mapping of pore occupancy and fluid distributions has been achieved by the development of specialized MRI techniques [100, 101].

MRI enables direct, non-invasive observation of gases and fluids, including methane and $CO_2$ mixtures within porous media. Methane transport and $CH_4$-$CO_2$ exchange in sedimentary gas hydrates phase behavior, with signals detected in fractures are shown to increase with $CO_2$ exposure time [102]. Phase differentiation can be enhanced using contrast agents or isotopic substitution (e.g., replacing $H_2O$ with $D_2O$) [103]. However, the presence paramagnetic minerals are often impractical in rocks, voxel resolution is generally insufficient to resolve pore-scale dynamics, and imaging performance is reduced in samples containing ferromagnetic minerals. Despite these limitations, MRI remains one of the few techniques capable of providing time-resolved, non-invasive visualization of fluid flow and recovery mechanisms in porous media [104].

#### *Positron emission tomography*

Positron emission tomography (**PET)** is based on the detection of positron-emitting radionuclides. When a positron emitted from a decaying nucleus annihilates with an electron, two 511 keV photons are produced in opposite directions. A surrounding detector array registers these photons in coincidence, and the signals are processed to reconstruct the 3D spatial distribution of the tracer-labeled phase. PET's spatial resolution is fundamentally limited by the positron range and photon detection physics, while detector geometry, sensitivity, and signal processing also influence accuracy [105]. This technique is uniquely suited for non-invasive, *in situ* measurements of fluid flow and transport in porous media, as it can quantify tracer concentrations regardless of fluid density or optical transparency [106-108]. PET allows time-resolved imaging, enabling the study of dynamic displacement, mixing, and transport phenomena at both laboratory and field scales [106-108]. Its high-energy γ-rays penetrate confinement vessels at elevated pressures and temperatures, making PET suitable for reservoir-relevant conditions.

Recent studies have demonstrated the benefit of combining PET with X-ray CT [107, 109]. PET provides quantitative measurements of tracer-labeled fluid phases that are independent of density contrasts or rock structure, whereas CT yields detailed information on pore geometry and phase distribution when density differences are sufficient. Together, PET/CT imaging enables time-resolved (4D) visualization of displacement processes, fluid saturation changes, and the coupling between flow dynamics and rock heterogeneity. Compared with standalone CT, PET is particularly advantageous in low-porosity or low-contrast systems, where it provides more reliable saturation quantification and improved signal-to-noise ratios. Similarly, combining PET with MRI provides a powerful approach for studying multiphase flow in porous media [110]. This integrated method allows simultaneous, quantitative imaging of multiple fluid phases within the same system. By providing independent and complementary measurements, PET and MRI capture 3D fluid distributions with high reproducibility, facilitating detailed analysis of spatial and temporal flow dynamics.

### *3.3 Evaluation of three-dimensional imaging techniques*

3D imaging techniques provide volumetric insights into porous media, enabling quantitative analysis of pore geometry, connectivity, fluid distributions, and particle transport. **Table 2** summarizes the main 3D imaging methods, detailing their strengths, limitations, destructive potential, and typical combinations, highlighting how these complementary techniques span multiple spatial and temporal scales.

**Table 2.** Summary of three-dimensional imaging techniques.

| Imaging technique | Strengths | Weaknesses | Destructive | Often combined with |
|---|---|---|---|---|
| CT | Large sample volumes; 3D pore architecture, porosity, connectivity; relative permeability, capillary pressure, interfacial phenomena | Spatial resolution limited to micrometer scale; trade-off between field of view and voxel size; acquisition time | No | PET, MRI, FIB-SEM, CLSM |
| CLSM | High-resolution 3D imaging; non-destructive; | Limited penetration depth (<500 μm); small | No | Microfluidic devices, glass models, μCT |
| | visualization of biofilms, microbial processes, multiphase fluids; fluorescent labeling enhances phase contrast | volumes; requires transparent or thin samples | | |
| FIB-SEM | Nanometer-scale 3D resolution; pore network, grain arrangement, and elemental mapping via EDS; ideal for detailed structural analysis | Destructive; small sample volumes; extensive sample preparation; time-consuming | Yes | µCT, EDS, CLSM |
| MRI/NMR | Time-resolved imaging; visualizes fluids and gases (e.g., methane, hydrogen) in situ; sensitive to hydrogen-rich phases | Limited spatial resolution for pore-scale imaging; less effective with ferromagnetic minerals; voxel size may miss fine pore details | No | PET, µCT |
| PET | Time-resolved imaging; quantitative measurement of tracer-labeled fluids; effective in low-porosity/low-contrast systems; field- and lab-scale applications | Requires radioactive tracers; lower spatial resolution; safety and handling constraints | No | µCT, MRI, CT |

µCT and Neutron CT are non-destructive, reconstructed 3D imaging techniques that acquire volumetric data by combining multiple 2D slices (**Figure 2**). µCT offers high spatial resolution and excellent contrast for dense minerals, allowing quantitative characterization of pore architecture, porosity, connectivity, relative permeability, wettability, capillary pressure, and interfacial phenomena. Neutron CT, by comparison, provides strong contrasts for light elements and hydrogen-rich phases, making it particularly useful for studying fluid distributions in porous media. Together, these methods provide complementary structural and functional information across macroscopic and microscopic scales. Limitations include the trade-off between field of view and resolution, instrument access, and acquisition time.

CLSM and FIB-SEM enable high-resolution, nanoscale 3D imaging. CLSM generates 3D reconstructions from optical sections, allowing non-destructive visualization of microbial processes, biofilm formation, and multiphase fluid distributions in thin volumes, although its penetration depth is limited. FIB-SEM combines serial ion-beam milling with SEM imaging to reconstruct nanometer-scale pore networks and solid matrices. It provides elemental mapping via EDS, but is destructive, restricted to small sample volumes, and requires careful sample preparation.

Direct 3D imaging techniques capture volumetric information inherently, avoiding errors or distortions that can arise from image stacking and alignment in reconstructed datasets. Techniques like MRI and PET enable non-invasive monitoring of dynamic multiphase flow, solute transport, or gas migration in real time image (Figure 2). PET provides phase-specific information, such as tracer-labeled fluids, independent of optical transparency or density contrast. It is particularly advantageous in low-porosity or low-contrast systems and can be combined with CT to enable 4D visualization of displacement processes and fluid–rock interactions. MRI allows non-invasive visualization of fluids and gases, including methane and hydrogen mixtures, within porous media, capturing dynamic flow and transport processes without disturbing the system. Integrating PET with MRI or CT combines structural and tracer-based measurements, enhancing spatial resolution, phase distinction, and temporal tracking.

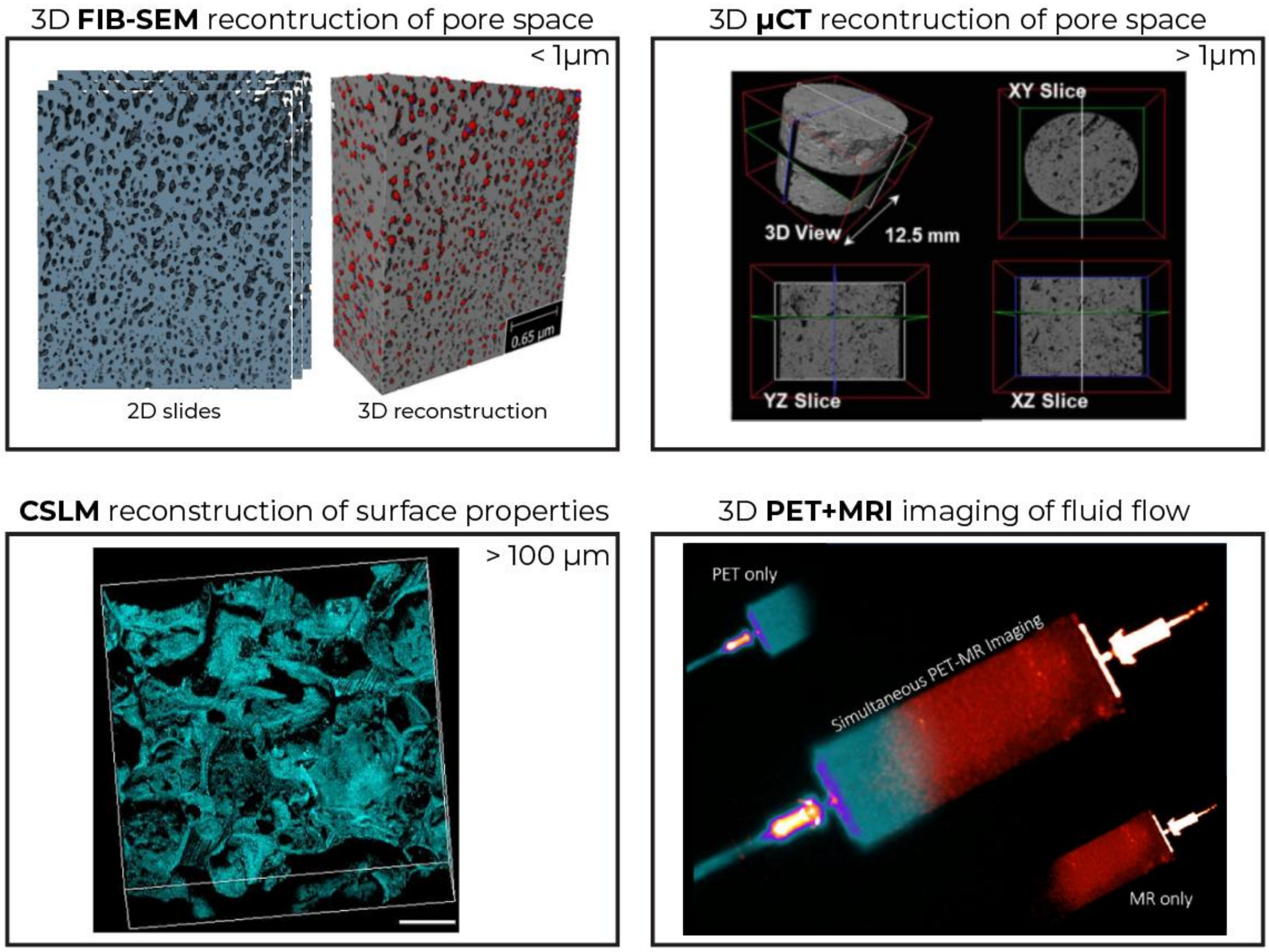

***Figure 2**. Examples of 3D images. Top-left: FIB-SEM 3D reconstruction of the pore structure in the asymmetric Viresolve® Pro virus removal filter. Reproduced with permission from ref [111]. Top-right: 3D image of a rock sample with three orthogonal slices through the volume. Reproduced with permission from ref [112]. Bottom-left: 3D reconstruction of porous limestone via CLSM. Reproduced with permission from ref [113]. Bottom-right: PET-MRI visualization of fluid dynamics within brine-saturated sandstone pores [110].*

## 4. Applications of imaging techniques in lab use

Laboratory imaging techniques provide critical insight into porous media behavior, allowing controlled experiments that link pore-scale processes to macroscopic phenomena. While Sections 2 and 3 focused on methodology, this section emphasizes applications, mechanistic understanding, and integrative approaches in laboratory research (**Table 3**).

**Table 3** Imaging techniques, their laboratory applications and length scale of observation.

| Methods | 2D/3D | Applications | Samples | Scales | | | | |
|---|---|---|---|---|---|---|---|---|
| | | | | nm | µm | mm | cm | m |
| Camera | 2D | Geometry, macroscopic features,Flow dynamics | TF, MC, TS | | | | | |
| OM | 2D | Pore structure down to ~200 nm, grain shape, | TF, MC, TS | | | | | |

| | | | | |
|---|---|---|---|---|
| | | mineral distribution, conductivity. | | |
| SEM | 2D | Static, high-res surface morphology (1–10 nm), pore throat structure, mineral-pore interfaces, micro-fractures. | TS | |
| TEM | 2D | Static, Nanopores (<10 nm), crystal lattice defects, pore connectivity at nanoscale. | TS | |
| AFM/STM | 2D | Static, nanoscale maps of pore walls, roughness, surface heterogeneity. | TS | |
| Raman/FTIR | 2D | Dynamics, chemical composition. | TF, MC, TS | |
| FIB-SEM | 2D→3D | Static, 3D nanoscale pore networks, connectivity, mineral–pore relationships. | TS | |
| CLSM | 2D→3D | Flow dynamics, 3D pore imaging with fluorescence, biofilms, tracer distribution. | TF, TS | |
| CT | 2D→3D | Flow dynamics, 3D pore architecture, porosity, permeability, multiphase flow. | TS, RC, CP | |
| NMR/MRI | 3D | Flow dynamics, in-situ flow imaging, saturation mapping, fluid distributions. | TS, RC, CP | |
| PET | 3D | Flow dynamics, tracer transport, flow pathways, mixing and dispersion. | RC, CP | |

TF: Transparent flow cells; MC: Microfluidic chips; TS: Thin section; RC: Rock cores; CP: Core plug.

### *4.1 Structural characterization of porous media*

Different imaging modalities serve distinct purposes depending on the scale and type of laboratory sample. SEM, TEM, and AFM are typically applied to thin sections, powders, or engineered porous materials to reveal surface morphology, mineral composition, and nanoscale textural features [5, 39]. These high-resolution methods are essential for identifying clay coatings, surface roughness, and reactive sites that strongly influence wettability and mineral–fluid interactions.

For intact rock cores and synthetic porous media, µCT provides non-destructive, 3D visualization of pore networks and fractures, enabling quantitative evaluation of porosity, connectivity, and volumetric distributions over representative volumes [1, 68, 72]. Time-lapse µCT further extends these applications by tracking dynamic microstructural changes during flow-through experiments, such as fracture propagation, mineral precipitation, or particle migration [19, 114].

OM, including polarized light and confocal imaging, remains a rapid and accessible approach for thin sections and transparent analog systems. It enables assessment of mineral textures, grain boundaries, and pore-scale heterogeneities, and when combined with fluorescence or confocal techniques, provides direct visualization of fluid distributions in synthetic porous media [18, 25].

### 4.2 *Multiphase flow and fluid dynamics in natural rocks*

Laboratory imaging of multiphase flow in natural rock samples provides critical insights into how fluids are distributed, displaced, and transported under controlled conditions that mimic reservoir environments. Intact rock cores, i.e., samples extracted from the subsurface without significant disturbance to their natural pore structures or mineralogical composition, are commonly used. They allow non-destructive and *in situ* observation of flow processes while preserving natural pore structures and mineralogical heterogeneity.

X-ray CT/μCT is one of the most widely applied modalities, enabling high-resolution 3D mapping of pore networks and fluid distributions in core samples. It provides quantitative characterization of connectivity, saturation patterns, and structural heterogeneity, and, when combined with time-lapse imaging, allows direct tracking of drainage and imbibition cycles, capillary trapping, and interface evolution [89, 115]. Complementary modalities such as MRI and PET are especially valuable for experiments involving multiple fluid phases, tracers, or compositional gradients. MRI enables non-invasive visualization of immiscible and partially miscible fluids within cores, while PET offers real-time quantification of dynamic transport and phase partitioning at the core scale [102, 109, 116]. These approaches facilitate the study of phenomena such as residual trapping, capillary fingering, and fluid–rock interactions under conditions closely resembling natural reservoirs.

While these techniques provide unique mechanistic insights, these techniques are resource intensive. High-resolution imaging often requires specialized equipment, long acquisition times, and carefully designed experimental protocols to ensure accurate representation of multiphase behavior [108]. Nevertheless, the integration of these methods has become essential for linking pore-scale processes to macroscopic flow behavior and for validating predictive models of multiphase transport in complex porous media.

### 4.3 *Synthetic geometries*

In the laboratory, complementary approaches using synthetic geometries with transparent windows are widely employed to facilitate camera-based visualization and microscopy-based imaging of fluid dynamics. These setups, which can range from microscale pore networks to larger room-scale models, allow controlled experiments that reveal fundamental flow and transport processes. Combining high-resolution imaging with such experimental systems capture fluid behavior across multiple length scales and may be used as input to upscale average displacement mechanisms from core samples to reservoir-scale applications (**Figure 3**).

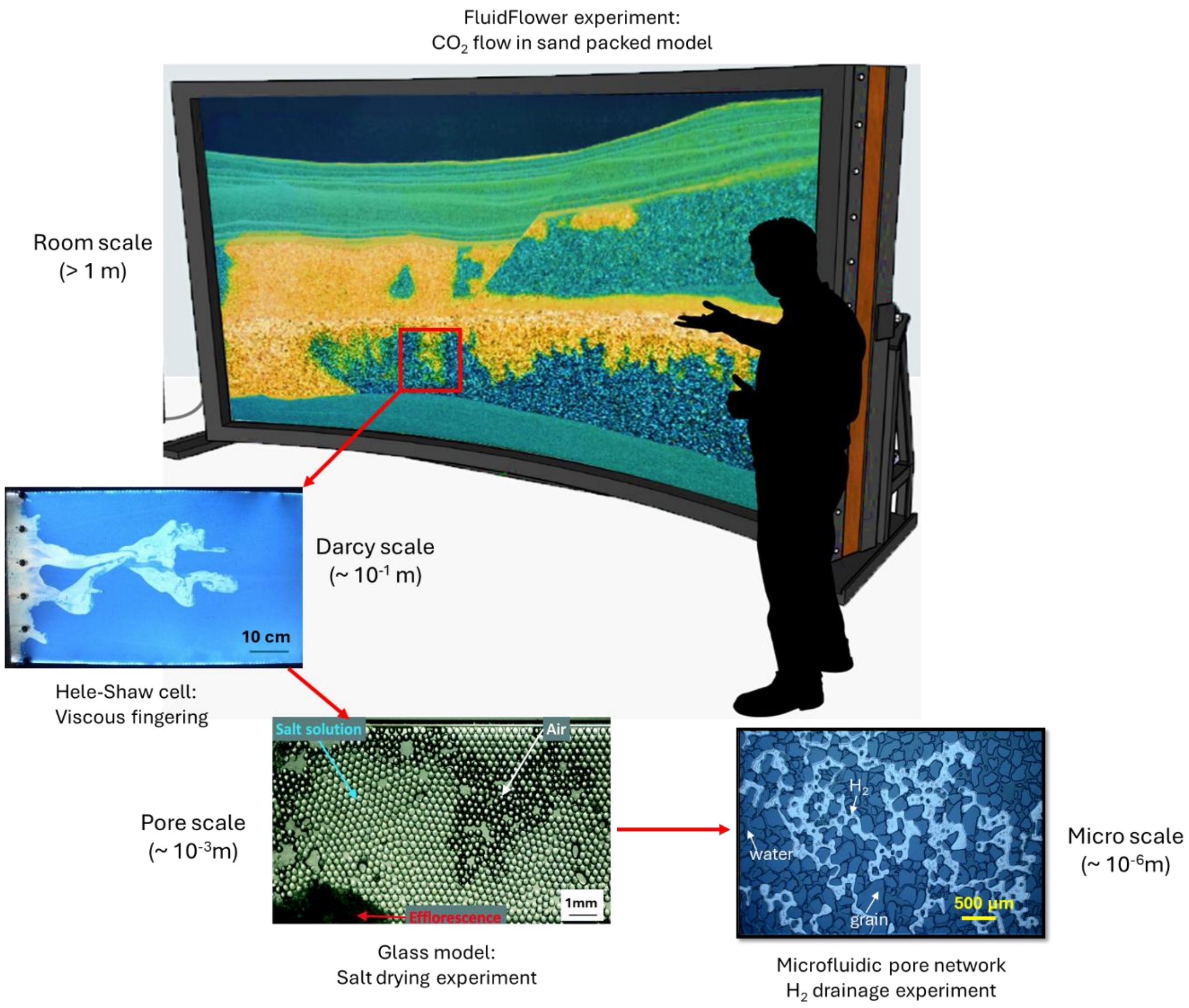

**Figure 3**. Multiscale lab methodologies developed to quantify multiphase flow and physicochemical interactions from micro scale (µm) to room scale (m). The Hele–Shaw cell image is adapted from [117] with permission. The glass model image is adapted from NaCl crystallization experiments in glass models [118].

### *Microfluidics/micromodel*

Microfluidic pore networks, commonly referred to as micromodels, are artificial, quasi–2D porous media designed to enable direct visualization of complex flow environments at the micrometer to millimeter scale. In recent years, micromodels have become indispensable tools for investigating multiphase fluid flow, pore-scale displacement mechanisms, and fluid–rock–gas interactions relevant to subsurface reservoirs. Advances in microfabrication, including etching, soft lithography, and photolithography, enable accurate replication of rock pore geometries in micromodels made from polymers, glass, or silicon. Polymers are low-cost and easy to fabricate but unsuitable for reservoir conditions, whereas glass and silicon are more robust, though all micromodels differ from natural rocks in surface chemistry, wettability, and mechanical properties. Photoelasticity can be applied to micromodels to visualize stress distributions and fluid–structure interactions during multiphase flow, providing insight into pressure-driven deformations and capillary stresses within pore networks [119]. More recently, mineral microfluidics—such as silicon micromodels with deposited calcite [120, 121] and real-rock micromodels [122], incorporating thin reservoir rock sections, have been developed to more accurately reproduce reservoir conditions and enable direct observation of pore-scale processes, including microbial growth and redox reactions.

The transparency of microfluidic devices allows real-time imaging of pore-scale processes with OM and CCD camera, providing insights into multiphase displacement, interfacial dynamics, wettability, capillary forces, and phase distribution, often complemented by particle image velocimetry (**PIV**) to map velocity fields (**Table 4**).

**Table 4** Summary of pore-scale observations and measurements across key properties.

| Properties | Qualitative Observations | Quantitative Measurements |
|---|---|---|
| Wettability & Interfacial Tension | - Visual tracking of droplet/bubble coarsening (e.g., Ostwald ripening) [123]<br>- Wettability state (hydrophilic/hydrophobic) via fluid adhesion [8, 124]<br>- Fluid-film coating on pore walls [19] | - Contact angle [8]<br>- Interfacial tension (IFT) from meniscus curvature [125]<br>- Droplet size distribution shift [126] |
| Pore Geometry & Transport Properties | - Pore-throat connectivity and network heterogeneity [19]<br>- Dominant flow pathways (e.g., preferential channels) [127] | - Permeability [128]<br>- Tortuosity index [129]<br>- Pore connectivity metric [130] |
| Rock–Fluid-Gas Interactions | - Mineral dissolution and precipitation patterns [19, 53]<br>- Fines migration and pore clogging [58] | - Reaction rates (e.g., $\mu m^2$/sec dissolution) [19]<br>- Permeability reduction (% change) [58, 124] |
| Flow Dynamics & Displacement Mechanisms | - Flow patterns: fingering, channeling, piston-like displacement, laminar and turbulent flow [125, 128-130]<br>- Snap-off events and Haines jumps (during drainage/imbibition) [131, 132] | - Capillary pressure–saturation curves [133, 134]<br>- Relative permeability vs. saturation |
| Pore-Scale Fluid Distribution | - Spatial arrangement of fluids (liquid/gas) [125, 128-130]<br>- Visual identification of trapped phases [123] | - Phase saturation (pore space coverage) [53, 124, 135]<br>- Cluster size distribution of residual fluids [8] |

### *2D visual glass model*

Unlike microfluidic chips, the glass model employs an assembly of transparent glass beads arranged in a thin, 2D plane to replicate the porous structure of geological formations at centimeter to meter scale [136]. The packed glass beads form interconnected pore spaces that mimic natural permeability and porosity, thereby serving as a physically realistic proxy porous medium. Brine or other reservoir-relevant fluids can be injected into the model to investigate multiphase flow dynamics, including displacement efficiency, fluid front propagation, trapping mechanisms, and flow functions [137]. The transparency of the glass and the 2D configuration allow direct visual observation and high-resolution imaging of interfaces, flow patterns, and pore-scale displacement processes [118].

Glass bead models are particularly useful for validating computational models and investigating fundamental processes such as capillary forces and the influence of pore-scale heterogeneity. The setup allows controlled experiments under varying injection rates, fluid compositions, and pressure gradients, making it widely applied in reservoir engineering, enhanced oil recovery, and hydrogeology [138]. In summary, the 2D visual glass bead model uniquely enables the observation of gravity-driven fingering in addition to pore-scale fluid displacement, offering a rare opportunity to study flow phenomena that depend on buoyancy effects in porous media [139]. Despite these advantages, glass bead models have important limitations. Their surfaces differ significantly from natural minerals in terms of chemistry, wettability, and reactivity, which restricts their applicability for studying biogeochemical processes or microbial activity, as microbial adhesion and growth are generally unfavorable on glass bead surfaces [140]. In addition, pressure conditions represent a major constraint: most experiments are conducted at ambient pressure, and the maximum reported operating pressure is only in the order of a few bars, well below reservoir conditions. Moreover, the spherical shape of glass beads differs from natural rock grains, which affects pore-scale flow properties, residual trapping, and capillary behavior, further limiting the representativeness of glass bead systems for reservoir-relevant studies.

#### *Hele-Shaw flow cell*

A Hele-Shaw cell is a laboratory device consisting of two closely spaced parallel plates, typically made of glass or transparent plastic, with a narrow gap that forces fluids to flow in a quasi-2D manner. Originally developed by Henry Selby Hele-Shaw, this setup is widely used to model flow in porous media by approximating Darcy flow conditions. Hele-Shaw cells are particularly valuable in studies of $CO_2$ sequestration, enabling visualization of density-driven convective mixing and the onset of Rayleigh instabilities during $CO_2$ dissolution in brine [141, 142]. The transparent structure allows the use of dyes, high-resolution imaging, and optical microscopy, making Hele-Shaw cells ideal for detailed experimental studies of multiphase flow and transport phenomena [117, 141, 142]. However, the quasi-2D geometry simplifies the 3D complexity and heterogeneity of natural porous media, which can influence flow patterns, fingering, and displacement efficiency. The uniform gap between plates does not capture the variability in pore sizes and connectivity found in real rocks, limiting direct extrapolation of results to subsurface systems [143]. Thus, while Hele-Shaw cells are excellent for investigating fundamental flow instabilities and multiphase dynamics, they cannot fully reproduce the behavior of fluids in natural porous media.

#### *FluidFlower*

Intermediate-scale (decimeter to meter) quasi-2D laboratory experiments are widely used to study multiphase porous media flow, including gravity unstable flows in the presence of heterogeneity [144-146] and $CO_2$ migration and dissolution [147-149]. These approaches enable visualizing and studying a range of porous media flow dynamics in engineered representative porous media. The FluidFlower experimental facility at the University of Bergen provides a versatile platform for studying meter-scale, quasi-2D multiphase flow in model geological geometries with high-resolution data acquisition [4, 10]. The setup consists of packed sands that replicate heterogeneous pore networks, allowing repeatable experiments without repacking between runs. Although operated at ambient pressure and temperature, the rig captures key subsurface processes relevant to $CO_2$ storage, including structural trapping beneath sealing layers, residual trapping in partially saturated zones, dissolution trapping when $CO_2$ dissolves into the water phase, and convective mixing leading to gravity-driven fingering [3]. Structural trapping in the rig is driven primarily by capillary entry pressures rather than permeability contrasts typical of field-scale systems, highlighting the need to carefully interpret scaling effects when extrapolating observations to real reservoirs.

### *4.4 Combined and multimodal approaches*

A growing trend in laboratory studies is the integration of multiple imaging and analytical modalities to capture complementary information across scales and processes. Microfluidic devices are particularly suited for multimodal integration. When coupled with Raman spectroscopy, they allow simultaneous visualization of pore-scale flow and *in situ* chemical reactions, such as mineral dissolution, precipitation, or reactive transport [53, 54]. CCD camera and OM can record multiphase displacement and interfacial dynamics in real time [13], while PIV maps local velocity fields within the pore network [24]. Photoelasticity applied to microfluidic or Hele-Shaw setups enables visualization of stress distributions and fluid–structure interactions, revealing pressure-induced deformations and capillary forces that influence pore-scale flow [119].

At larger scales, PET combined with MRI or X-ray CT enables 4D imaging of fluid transport in rock cores, providing both structural and tracer-based quantitative data [108, 110]. For nanoscale characterization, SEM applied to thin sections resolves surface features at nanometer resolution, offering detailed information on surface roughness, mineral heterogeneity, and reactive interfaces. These surface insights complement μCT measurements, which capture the 3D pore structure of the bulk sample, bridging scales from the macroscopic pore network to nanoscale surfaces [1, 90]. These integrative approaches are particularly valuable for validating models, improving mechanistic understanding, and providing datasets for upscaling laboratory observations to reservoir-relevant conditions.

## 5. Data analysis and availability

Image processing and data analysis are essential components of imaging workflows, serving as gateways to quantitative research. Except for a few formats, such as PET images, which directly provide inherently quantitative signals, these processes are crucial for extracting meaningful, quantitative information from images. However, limited access to metadata and raw data can significantly reduce the potential for reuse, especially in interdisciplinary fields like porous media research. As interest grows in high-resolution datasets for data-driven modeling and the validation of traditional mathematical models, the need to follow the FAIR principles (Findable, Accessible, Interoperable, and Reusable) [150] becomes increasingly urgent. In this context, we provide a brief overview of commonly used image processing and data analysis tools, trends, and assess the current state of open-access data availability in porous media research.

### *5.1 Quantitative data extraction from images*

A central aspect of any image-based analysis is the scale of interest. While the respective imaging technology sets the upper limits of resolution and detail, the nature of the research question often determines the appropriate scale for analysis. For example, studies aimed at validating continuum-scale models require compatible continuum-scale data, which may necessitate some form of upscaling given pore-scale data; or on the contrary resolution enhancement is envisioned effectively taking the role of downscaling. Concrete examples include buoyancy-driven mixing at the Darcy scale both in homogeneous porous media [151], or in heterogeneous systems where spatial variations in capillary pressure may play a significant role requiring facies-conforming upscaling [10]. The combination of image representation at disparate scales - such as μCT and SEM - opens ways for multi-scale image analysis [89, 152]. In our survey of the references cited in this review, fewer than 10% of studies employ such multi-technique approaches, while the majority continue to focus on single-scale imaging.

Upscaling is inherent to porous media research. It requires the detection of the pore space if resolved, typically through segmentation. Additionally, the effective conversion from pore-scale to Darcy-scale information requires classical volumetric averaging using a representative volumetric element [153], variational smoothing [154] or other suitable techniques. With an immense choice of traditional or deep learning smoothing algorithms, there is an entire plethora for converting data to larger scale, while potentially taking account for structural details as multi-scale heterogeneities. However, upscaling techniques need not be equivalent nor interchangeable, and both quantitative and qualitative results may depend on the choice of upscaling methodology [155, 156]. Downscaling in the form of image resolution enhancements have recently seen a new advent. The use of deep learning and combination of large-scale low-resolution and patches of high-resolution images has enabled resolution enhancements of static pore structures [157] and dynamic fluid flow [158]. Since the temporal resolution is dictated by the imaging method, super-resolution techniques open new ways of extracting high-resolution information from low-resolution images. Combining multiple low-resolution snapshots with a few high-resolution images reconstructs fine-scale features—fast multiphase flows, small heterogeneities, and pore-scale mixing—enhancing modeling and upscaling [152, 159].

If noise appears at the same scale as the scale of interest, noise removal is an integral component of any standard image processing workflow [160]. Traditional filtering [161] and total variation denoising [162] are common choices among many others developed in the wide field of image processing. However, machine learning has also reformed classical image processing tasks like noise removal [163, 164].

Image processing workflows are typically designed as sequential-in-time. For each snapshot in time, the spatial image is processed separately, involving upscaling, denoising, and further data analysis. We note that despite the immense advancements in imaging and generation of high-resolution spatio-temporal data, dedicated workflows for dynamic imaging remain scarce and instead time-resolved images are analyzed by stacking static-in-time analyses on top of each other. Indeed, it is straightforward to envision scenarios where the 4D methodology could be effectively applied [165, 166]. For instance, noise in imaging is typically non-static in space and sparse over time, making space-time algorithms particularly well-suited for efficient noise reduction.

Segmentation is arguably the most widely used technique to extract quantitative information from images, identifying clearly separable structures as pores and the volumetric composition of immiscible fluid phases with sharp interfaces [167]. The detected pore space facilitates pore scale simulations enabling statistical approaches for extracting effective parameters as porosity, permeability and tortuosity [168]. Trained deep learning models enable circumvention of costly simulations and direct correlation of pore space configurations and material properties [169]. Segmentation plays a big role in Digital Rock Physics and the automated generation of synthetic porous materials using machine learning [170]. To segment unimodal images, both value-based and gradient-based thresholding, e.g. Watershed algorithm [167], are frequently applied techniques. These require user-defined threshold values which are either manually calibrated against physical measurements [171] or utilizing automated algorithms like Otsu-thresholding [172]; by definition, segmentation builds on the assumption that fluid and solid phases appear immiscible and clearly distinguishable. With either time-varying light conditions, or physical or chemical processes altering the signal intensity, threshold values may need to be adapted in time to ensure structure-preserving phase detection. Moreover, segmentation uncertainty [173] as well as the arbitrariness of choosing a static threshold parameter when segmenting continuous data [174] showcase the potential limitations of segmentation.

Regression aims at the detection and transformation of continuous signals as opposed to segmentation. The use of regression is central in the analysis of miscible fluids when mixing occurs at higher resolution than the image resolution [175] or when the Darcy-scale is the scale of interest [176]. In such examples, identifying concentration and saturation gradients at the continuum scale defines the primary task. Image formats as PET enable direct interpretation of a continuous signal, while uni- or multi-modal photographs enable correlation between grayscale or colors and concentration and saturation values [177]. The signal intensity, however, is strongly determined by the conditions during image acquisition including uniform illumination, color temperature of the ambient light, and equipment as a lens, which together highlight the importance for robust routines for calibration as well as uncertainty estimation.

In a wider sense and similar to multi-scale imaging, multi-modal imaging does not only offer potential in improved visualization but also improved data analysis. Offering multiple views and data formats has the potential in increasing information content as well as reducing the uncertainty, e.g., in segmentation as so far mostly employed in medical imaging [178]. While measurement errors and signal-to-noise ratios are typically known, uncertainty quantification in the data analysis are often insufficiently investigated or documented as identified in a recent review on porosity estimation using thresholding [179]. Often, the uncertainty is estimated merely in terms of computational flow simulations [180], underestimating the overall uncertainty.

Free and open-source platforms, such as FIJI/ImageJ, are widely used for image segmentation, providing versatile tools for image processing and analysis [1, 8, 9]. The use of scripting and programming, particularly with MATLAB or Python libraries such as OpenCV and scikit-image, has become a widely adopted practice, resulting in the development of various specialized porous media toolboxes such as PoreSpy [181], DarSIA [154], PuMA [182], and DragonFly [183]. With some coding skills, these platforms enable customized workflows, access to extensive scientific libraries, and reproducible pipelines for large-scale data processing and quantitative visualization [19, 72, 123, 124]. Artificial intelligence and machine learning approaches are increasingly integrated into image analysis workflows to enhance resolution, segmentation accuracy, automated feature extraction, and enable high-throughput processing of large image datasets [52, 59, 159]. These approaches can be implemented in MATLAB or Python using widely available machine learning and deep learning libraries (e.g., TensorFlow [152], PyTorch [115], Unet [52]), allowing researchers to develop tailored pipelines for both 2D and 3D imaging data. In parallel, several commercial software packages have been developed for specific imaging modalities; for example, the Avizo Fire series is extensively used for processing and analyzing CT datasets [63, 65, 184].

### *5.2 Data availability assessment – a community perspective*

Experimental data holds a dual value: while typically generated to investigate specific properties or phenomena, image data in particular offers immense potential for cross-disciplinary reuse. In physics-based, data-driven, hybrid, and digital twin modeling, reliable reuse of datasets is crucial – whether for comparative validation or for training, learning and identifying physical behavior.

The importance of unified data formats and rich metadata cannot be overstated. When raw data is accompanied with detailed metadata and well-documented experimental protocols, it transforms raw data into a comprehensive reference framework, ideal for modeling, validation, and reproducibility. Clear documentation, whether explicit or implicit through image processing and/or data analysis codes, enables reusability and interoperability, which are critical for quantitative data comparisons e.g. against simulations.

Broader reuse is often limited by inconsistencies in data formats, incomplete metadata, and insufficient documentation of experimental setups. To ensure comparability, metadata should include clear identifiers of coordinate systems and reference points to allow for aligning geometries. It should also contain experimental conditions such as pressures, temperatures and boundary conditions, especially when not reported in the associated publication.

As a result of restricted open data availability, validation studies frequently present experimental data, simulations, and comparisons within the same publication [185], or at least by groups with full access and insight [10, 75, 104]. While practical for single pointed focused studies, this practice hinders large-scale comparative modeling, which typically require coordinated community efforts. These structural limitations represent missed opportunities, as open and accessible data are essential for reproducibility, cross-validation, and collaborative research.

Ensuring data availability is a first critical step toward adhering to the FAIR principles [150]. As already emphasized, this encompasses raw data, metadata, and ideally image processing scripts, which together provide quantitative data. Ultimately, broad adoption of open-access practices, comprehensive metadata documentation, and transparent reporting of workflows – including standardized imaging parameters, experimental conditions and data-processing workflows – accelerate scientific progress, enhances model generalization, and supports predictive frameworks for porous media research.

Examples aligned with these recommended best practices in data sharing – showcasing the generated potential for interdisciplinary collaboration – include a model verification study of tracer transport in fractured media based on PET imaging compared against simulations [185], and research on multiphase flow in complex arranged multi-layered sands conducted within the FluidFlower framework [3, 10]. Both examples employ the open-source image analysis toolbox

DarSIA which facilitates transparent and reproducible workflows allowing direct publication of image analysis scripts alongside datasets, e.g. through persistent *Zenodo* repositories [186]. Flexible in transforming images and data, DarSIA serves as a practical adapter between experimental and computational research, exemplified in benchmark initiatives assessing modeling capabilities of $CO_2$ storage [154] as well as digital twin research [2]. Ultimately, the developed digital twin project showcases the potential of cloud-integrated digital twins, coupling experimental rigs with simulators via real-time data sharing. The setup supports on-the-fly image processing, data-informed corrections of simulations through hybrid modeling, and feedback loops that steer injection protocols using optimal control strategies.

Complementary initiatives such as the Digital Rocks Portal [187] and the OpenPNM modeling platform [188] further demonstrate how openly available µCT datasets and pore-network workflows can provide common benchmarks for segmentation, pore-scale simulation, and model validation. Such frameworks highlight the transformative impact of open data and interoperable tools on reproducibility, cross-validation, and collaborative modeling at scale. While standardized data formats and open data sharing are envisioned to simplify interdisciplinary collaboration, these studies have nonetheless required close coordination between participating research groups to ensure comprehensive understanding of both experimental and simulation data.

To evaluate the current state of the research landscape with respect to data availability and its connection to published work, we conduct two complementary assessments. The first involves a manual review of the references cited in this study. For each cited work, we examine the availability of the publication (i.e., whether it is published with open access or not), the availability of data (whether it is provided via a persistent repository/hosted on a webpage, available upon request, or not accessible), and the availability of code, using the same criteria as for data. This assessment is intended to reveal temporal trends in research transparency and reproducibility.

The second assessment provides an automated assessment of the most recent publications submitted to *Transport in Porous Media* (TiPM) and *Computational Geosciences* (CompGeo). These journals serve experimental and computational research, making it suitable for comparing practices across these domains. By contrasting the author-selected references with a broader sample from the community, we aim to provide both a potentially biased individual perspective on influential papers and a more representative view of current practices within the field. At the same time, the community-based analysis serves as control group of the author-biased analysis.

### *View onto an author-biased selection: References cited within this work*

**Figure 4** presents the assessment of availability for papers, data, and code for the references cited in this work, where data corresponds to images and code corresponds to processing tools. The observed trends, particularly from 2005 onward, align with expectations: the research community has increasingly prioritized making their work accessible. While open access publication has become widespread, the sharing of data and code is lagging behind; 25-40% of the data is available (from openly to on request) and code being substantially less available. Nevertheless, a noticeable increase in the availability of these resources has emerged over the past 5–10 years, suggesting continued progress in the near future. While plotted together, we point out that the open access availability is not directly comparable to the other categories, since the published paper has always been properly archived. This has allowed older papers, which were originally published behind a paywall, to now retroactively be converted to open access.

One noteworthy pattern is the tendency to offer data and code only upon request. In terms of absolute numbers, this conditional sharing is comparable to full access. This practice poses a risk to long-term availability. Although such resources may appear accessible, their availability is often time-limited. As researchers change institutions, leave academia, or update contact information, access via personal request may effectively become unavailable. This highlights the importance of persistent and openly accessible repositories for ensuring reproducibility and transparency in research.

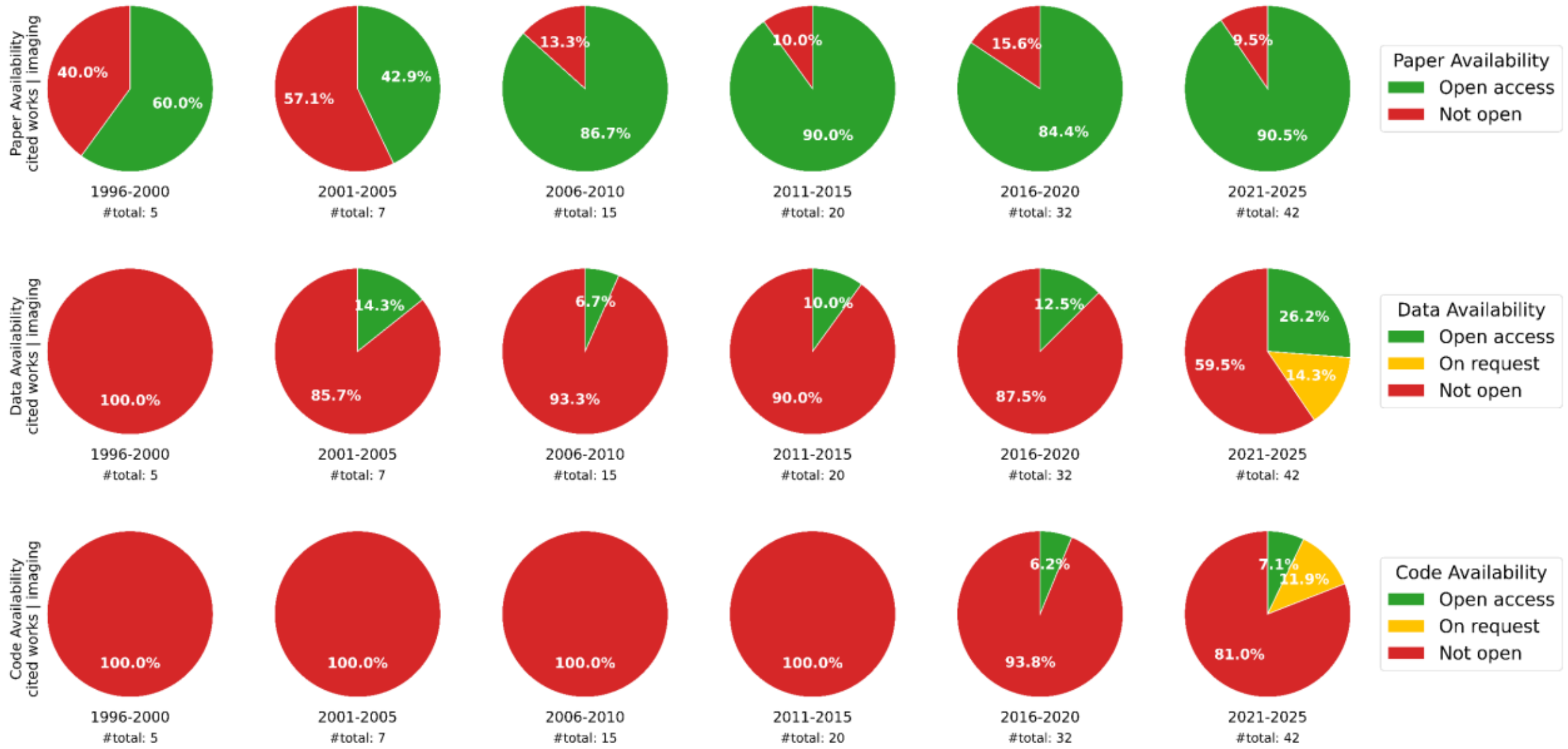

**Figure 4**. Trends in paper availability (row 1), data availability (row 2), and code availability (row 3) based on the references of this work (after 1995). The papers are grouped in 5-year periods and put in relative context within this period. The number of papers considered for the different periods are displayed below each respective pie chart.

*View onto the community: Submissions to Transport in Porous Media*

We pursue a similar objective as in the preceding analysis but extend our focus to assess the open access status of publications and the availability of associated data across the broader research community. To this end, we examine articles published in the cross-disciplinary but domain-specific journal TiPM. This enables an interesting comparison of different domains within porous media research. The computational domain has seemingly pushed open-source coding with various standardized computational platforms arising from research, but also various in-house code developments, made available through large hosting platforms as *GitHub* and national or institutional data repositories. We leverage the opportunity and put the availability of code (here also handled as data) in the computational domain in contrast to the availability of data in the experimental domain. By considering the same journal, which invites both experimental, computational and theoretical contributions to porous media, we eliminate also journal-specific preferences in making data available.

Using the journal's API, we retrieved the latest 1,000 entries in the TiPM database (as of 22 Nov 2025), covering the time frame of 2019-2025. Non-article entries (e.g., editorials, errata) were excluded based on metadata classification. The remaining articles were categorized by identifying keyword occurrences indicative of specific research themes, cf. Appendix B. We focused on two primary categories: *imaging* (identified through keywords such as 'tomography', 'synchrotron', etc.) and *simulation* (identified through keywords such as 'finite volume method', 'simulation', etc.), focusing on traditional simulation methodology. To further refine the classification, we also tagged articles related to *machine learning* (via 'PINN', 'ML', etc.) and *theoretical studies* (via 'linear stability', 'analytical solution', etc.), allowing us to proactively exclude works outside the scope of our target categories. As a result, almost 140 articles have been categorized as imaging articles, while almost 350 articles have been categorized as simulation articles. By leveraging the standardized article structure within TiPM, providing for instance the persistent sections *Rights and Permissions* and *Data Availability,* consistent metadata could be retrieved enabling automated classification of the paper and data availability. Similarly to the abstracts, these sections were compared against keywords uniquely associated to type of access, being *open access, on request* (if applicable)*, and no access*.

Since our approach is automated, it has several limitations that require addressing. First, relying solely on the abstract provides only a partial view of an article's content, which may lead to wrongful classification. Moreover, we acknowledge that the selection of keywords is biased by the authors' understanding of the different fields – noting however that the authors of this review have expertise across the considered domains. In cases where keywords from multiple categories are present, the keyword-based semantic evaluation has assigned article to the dominant category. As a result, interdisciplinary contributions, e.g., image-based machine learning, pore scale simulations based on µCT images, or model validation based on experimental data, are assigned to a single domain. Similarly, theoretical modeling studies involving simulations which are not explicitly framed in the abstracts are overlooked and ignored. The algorithm accounts for data availability statements found in other sections, such as "Acknowledgements" among others. However, it overlooks cases where information on data sharing is provided within the main body of the article (e.g. a remark in a methodology section or a footnote) rather than using the standardized "Data availability" section required by the TiPM style. Furthermore, we do not differentiate between types of code, such as image processing scripts versus full simulation frameworks, which limits the granularity of our analysis as compared to the above manual assessment. Finally, we do not verify whether the referenced data is genuinely accessible or reusable; our assessment is restricted to the mere presence of data availability statements and persistent identifiers.

In light of the aforementioned uncertainties, rather than aiming for precise metrics, our focus lies primarily on identifying broader trends and facilitating qualitative comparisons across the scrutinized research domains. This includes comparisons with the previously discussed analysis of cited works, providing context for the observed patterns. To estimate the uncertainty of the analysis, we perform a statistical robustness check. We randomly pick 10 articles from each category (*imaging*, *simulation*, *other*) for each data availability type (*open access*, *on request*, *no access*) resulting in total 90 out of 1,000 articles. For these, we check the assignment of the algorithm. Less than 5% of the tests were negative. Most negative results appear when checking articles in the *other* category, which for instance have been categorized as theoretical instead of computational. In contrast, *imaging* and *simulation* articles are safely identified, with the exception of interdisciplinary works ending up in both categories. While focusing on overall trends, we deem the uncertainty of the classification acceptable.

**Figure 5** displays the assessment of paper and data availability within the domains of imaging and simulation, based on the database curated from TiPM submissions. In contrast to the above manual analysis of the herein cited works, a lower paper availability can be identified within the community. Whether influential papers, considered in this work, had an impact just because of their paper or data availability is not necessarily justified; instead, the authors appoint their bias toward the perceived importance of methodological novelty, domain relevance, or community alignment. As expected, the availability of the publication texts is increasing over time, with increasing demand or request by funding agencies. Interestingly, however, the accounts of data availability are quite similar to the manual assessment. Increasing in time with a strong rise in recent years, data sharing practices are becoming more prevalent. But again, data access merely on author contact and on request are equally present as open access. When comparing publications within the two imaging and simulation categories, no categorical difference can be observed. The trends in both paper and data availability evolve very similarly, despite the much more present push for open-source code development in the computational domain. We, however, emphasize again, that the presence of data alone especially in experimental domains is often not sufficient for fully reproducing data or reuse these in cross-disciplinary contexts.

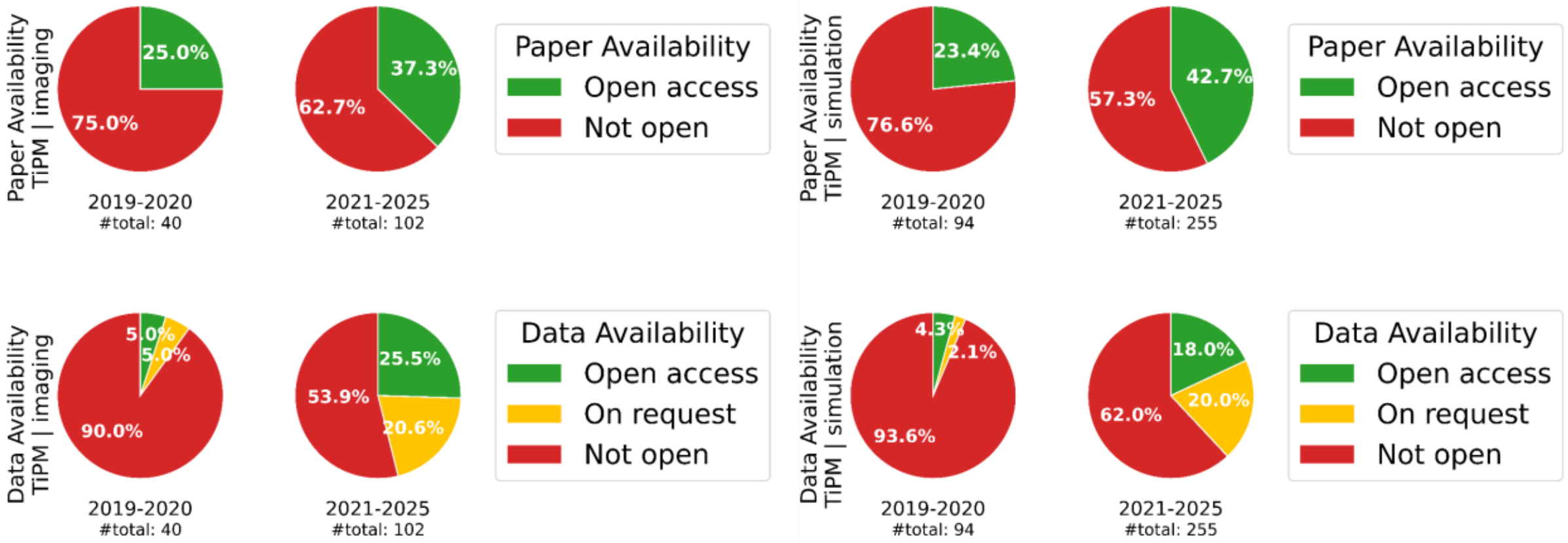

**Figure 5**. Trends in paper availability (row 1) and data availability (row 2) for the different categories of “imaging” and “simulation” papers published in TiPM within the time frame 2019-2025. The papers are grouped in 5-year periods and placed in relative context within this period. The number of papers considered for the different periods are displayed below each respective pie chart.

### *View onto the community: Submissions to Computational Geosciences*

To add a second, unbiased view onto the computational domain supporting the above numbers, we additionally consider the journal *Computational Geosciences*. Adhering to the same standardized format as TiPM since 2021, the unified analysis can be re-applied to obtain another control group representing the wider porous media community. Despite a predominance of computational articles, there also exist a few articles including experimental and imaging studies – typically interdisciplinary studies combining simulations and imaging or image analysis works. Leveraging the official API, one can retrieve the latest 1,000 articles published in CompGeo (as of 22 Nov 2025). Due to change in format and style of the published articles over time, we restrict the scope of the data availability assessment to articles from 2021-2025, reducing the database to 344 articles with ensured compatible format – paper availability can be still reliably assessed for all 1,000 articles. For the entire database, 30 articles have been identified as *imaging* articles suggesting little statistical significance, while 355 articles mentioned selected *simulation*-related keywords in their abstract – a statistical robustness check confirms the same reliability of the automated categorization as for TiPM. **Figure 6** displays the resulting availability of papers and data across the different categories. The results are overall in line with the analysis of the herein cited works and TiPM, again supporting the recent commitment towards increased data sharing. A slightly higher account of data availability ‘upon request’ is observed.

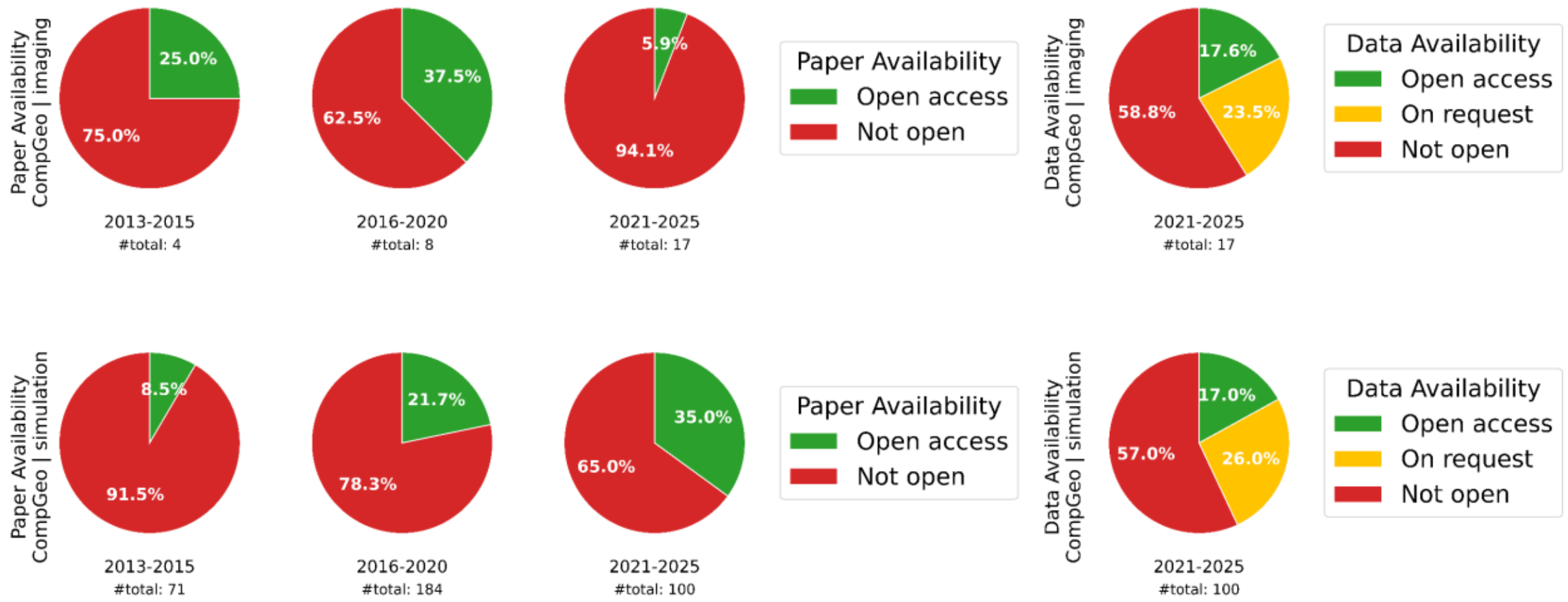

**Figure 6**. Trends in paper availability within 2013-2025 (column 1) and data availability within 2021-2025 (column 2) for the different categories of "imaging" and "simulation" papers published in the journal *Computational Geosciences*. The papers are grouped in 5-year periods and placed in relative context within this period. The total counts of papers considered for the different periods are displayed below each respective pie chart.

### *Concluding on data availability: recent developments*

The above three-fold analysis provides an unambiguous trend, in particular with the community-wide assessments supporting the manual assessment of the cited works herein. Practices for data availability are on the rise, while open access availability of publication texts has been a common practice for a longer while, yet still rising. No significant differences in practices were observed between the computational and experimental domains.

We conduct a final, slightly more refined assessment of the TiPM and CompGeo submissions. Using a similar keyword-assisted categorization, we group the articles into two overarching categories: *experimental/computational* (for which paper and data/code availability is expected) and *other* (e.g. theoretical works for which data availability is not expected). The respective trends for the experimental/computational works are displayed in **Figure 7**. The analysis resolves the above drastic increase of data/code availability and showcases a rather young and monotone increase in data sharing. With a short time lag after the introduction of FAIR in 2016 [150], increasing data sharing can be observed across the disciplines within porous media research. While there are no signs of a break in the trend, it is noteworthy to see the wide sharing of data 'on request'. The trend for this type of sharing develops equally strongly as open sharing.

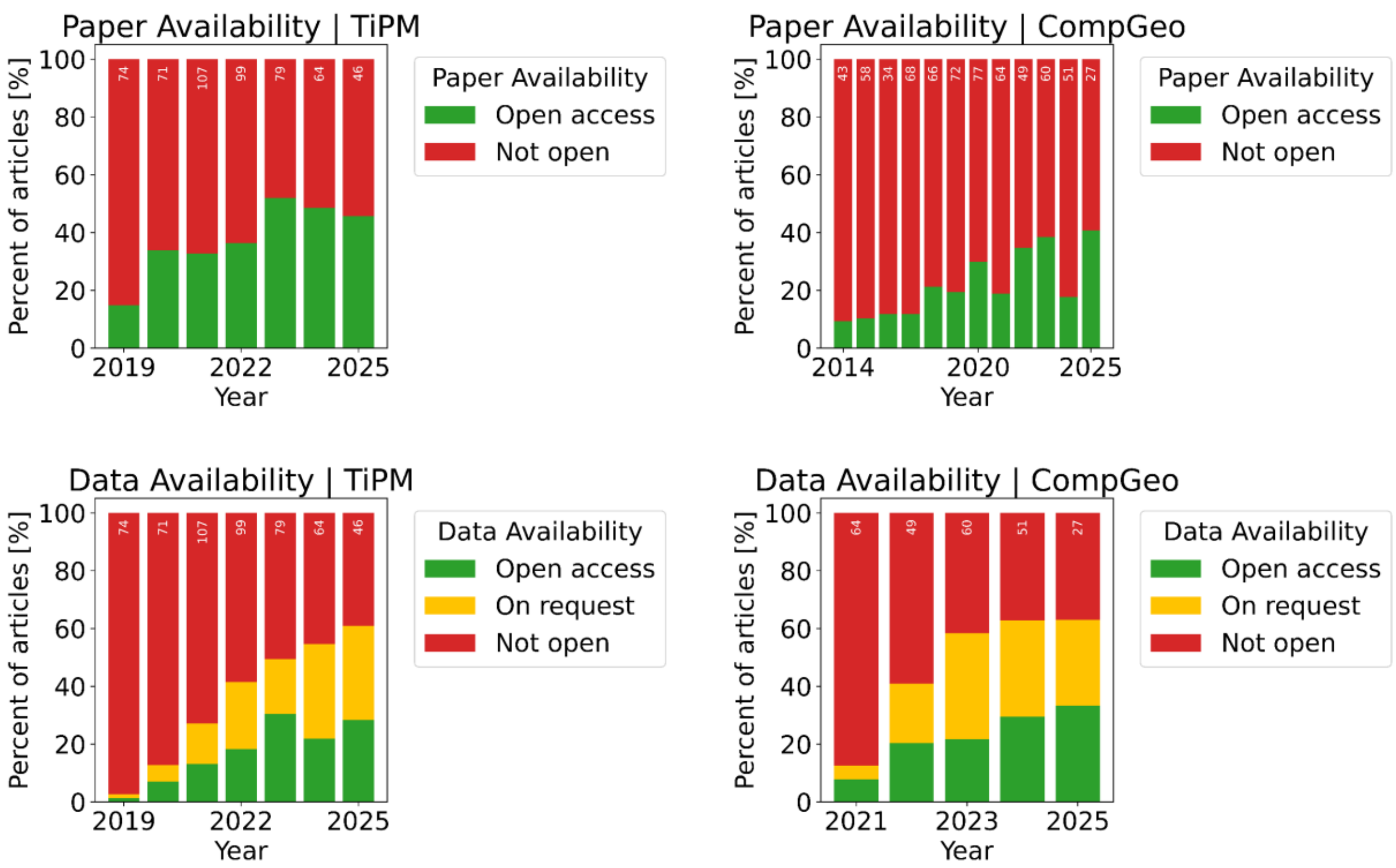

**Figure 7**. Relative trends in paper (row 1) and code/data availability (row 2) across experimental and computational domains combined in TiPM (left) and CompGeo (right) among their latest publications, respectively (restricted in the bottom right due to format). Absolute number of computational and experimental articles taken into account displayed on top of each bar.

## 6. Perspectives and Conclusions

Laboratory-based multiscale imaging has transformed our ability to observe, quantify, and interpret fluid–solid–biological interactions in porous media. By bridging pore-scale processes to macroscopic behavior, these methods provide the mechanistic foundation for validating modelling and validation technology within energy production/storage, $CO_2$ sequestration, groundwater management, and contaminant remediation. The integration of diverse 2D and 3D modalities has enabled unprecedented insights, yet challenges remain in scaling laboratory observations to field conditions, ensuring data quality, and fostering open data practices. Advances in image processing, segmentation, regression, and machine learning have enhanced resolution, automated feature extraction, and high-throughput analysis. Nevertheless, challenges remain in upscaling, segmentation uncertainty, and dynamic workflow analysis. In parallel, ensuring FAIR-compliant and openly accessible datasets is critical for reproducibility, cross-disciplinary integration, and large-scale comparative modeling. Community-wide assessments indicate a steady increase in open data and code availability, though a substantial fraction of datasets is still shared only 'upon request', which raises concerns about their long-term availability. Continued adoption of standardized metadata, persistent repositories, and transparent workflows is therefore essential to fully leverage multiscale imaging for predictive porous media research and to enable collaborative, data-driven advances across disciplines.

Future progress will depend on advances in both experimental methods and data-driven approaches. On the experimental side, the development of higher resolution, faster, and multimodal imaging platforms will allow researchers to capture transient processes with improved fidelity. On the data side, embracing best practices related to open data, shared repositories, and standardized reporting, is crucial to ensure that research results are reproducible, and enhance cross-laboratory comparability and accelerate innovation through interdisciplinarity and large data technologies.

A particularly transformative frontier lies in the application of Artificial Intelligence (AI) to porous media research. Within this broader field, machine learning and deep learning are already accelerating image analysis by improving segmentation, classification, and feature extraction, while reducing manual intervention and bias. They also support the generation of synthetic datasets, noise reduction, and quantitative mapping of flow and transport from 2D and 3D images. Beyond these, AI encompasses integrative frameworks that combine imaging, modeling, and decision-support tools — for example, digital twins that dynamically reconcile laboratory observations with simulations across scales. Such integrated lab-to-simulation workflows will open new possibilities in experimental design and execution. Looking forward, the synergy between advanced imaging, numerical modeling, and AI promises to reshape laboratory porous media research. To fully realize this potential, the community must focus on refining algorithms, democratizing access to open-source platforms, and building collaborative research networks.

## CRediT authorship contribution statement

Na Liu: Conceptualization, Data curation, Investigation, Writing – original draft; Jakub Wiktor Both: Data curation, Investigation, Funding acquisition, Writing – original draft; Geir Ersland: Methodology, Writing – review and editing; Jan Martin Nordbotten: Supervision, Funding acquisition, Writing-editing; Martin A. Fernø: Supervision, Data curation, Funding acquisition, Writing – original draft.

## Declaration of Competing Interest

The authors declare no conflict of interest.

## Acknowledgements

The authors would like to acknowledge the support provided by University of Bergen for supervision and resources.

## Funding sources

This work was supported by the HyDRA project – Diagnostic Tools and Risk Protocols to Accelerate Underground Hydrogen Storage (project no. 101192337), co-funded by the European Union through the Clean Hydrogen Partnership and its members; and by the Research Council of Norway under the projects Microbiological Opportunities and Challenges of Hydrogen Underground Storage (project no. 344183), Centre for Sustainable Subsurface Resources (project no. 331841), and Unlocking maximal geological $CO_2$ storage through experimentally validated mathematical modeling of dissolution and convective mixing (project no. 355188).

## Data availability

The generated database, the associated scripts, and the analyses of TiPM and CompGeo in Section 5 are available online at https://github.com/pmgbergen/trends_in_porous_media_laboratory_imaging_and_open_science_practices.

## Appendices

### *A. List of Abbreviations*

| Abbreviation | Full Term |
|---|---|
| **2D** | Two-Dimensional |
| **3D** | Three-Dimensional |
| **AFM** | Atomic Force Microscopy |
| **AI** | Artificial Intelligence |
| **BSE-SEM** | Backscattered Electron Scanning Electron Microscopy |
| **CCD** | Charge-coupled device |
| **CLSM** | Confocal Laser Scanning Microscopy |
| **CompGeo** | Computational Geosciences |
| **CT** | Computed Tomography |
| **EDS** | Energy Dispersive X-ray Spectroscopy |
| **ESEM** | Environmental Scanning Electron Microscopy |
| **FTIR** | Fourier Transform Infrared |

| Abbreviation | Full Term |
|---|---|
| **FIB-SEM** | Focused Ion Beam Scanning Electron Microscopy |
| **FM** | Fluorescence Microscopy |
| **NMR** | Nuclear Magnetic Resonance |
| **MRI** | Magnetic Resonance Imaging |
| **μCT** | Micro-Computed Tomography |
| **NMR** | Nuclear Magnetic Resonance |
| **OM** | Optical Microscopy |
| **PET** | Positron Emission Tomography |
| **PIV** | Particle Image Velocimetry |
| **PLM** | Polarized Light Microscopy |
| **SEM** | Scanning Electron Microscopy |
| **STM** | Scanning Tunneling Microscopy |
| **TEM** | Transmission Electron Microscopy |
| **TiPM** | Transport in Porous Media |

### *B. Categories and keywords*

The analysis in Section 5 associates keywords with categories. The association is automated and the details are found at the enclosed open-source repository (see Section Data Availability). For comprehensiveness, we list the keywords for the classification of imaging, simulation and other categories here.

| Category | Keywords |
|---|---|
| Imaging | Image, imaging, SEM, tomography, MRI, NMR, resonance, PET, CT, micro-CT, positron, xray, X-ray, microscopy, microscope, photo, spectromet, scopy, spectroscopy, raman, infrared, synchrotron, laser, segmentation, denoising, fiji, imagej, opencv, scikit-image, dicom, avizo, dragonfly, geoslicer, puma, pypore, sequencing, image analysis, porespy, glass beads, visualized, digital rock, microfluidic |
| Simulation | Numeric, spe10, mrst, COMSOL, open foam, openfoam, simulation, large eddy, LES, data assimilation, mesh, discrete element, discrete-element, finite element, finite-element, fem, finite volume, finite-volume, FVM, finite difference, finite-difference, FD, lattice Boltzmann, molecular dynamics, partial differential, monte carlo, simulator, simulated, history matching, pore network, pore-network, dns, lattice-boltzmann |
| Other | Solver, stability analysis, linear stability, linear instability, analytical solution, convergence, computation, density function, neural network, PINN, CNN, GAN, |
| | ANN, AI, deep learning, machine learning, laboratory, function, fractional flow theory, homogenization, boundary value, problem, solution, mathematical, centrifuge, review, constitutive, microfluidic, algorithm, theoretical, analytical, estimate, seismic, gaussian |